\documentclass[fleqn,twoside]{article}
\usepackage{espcrc2}

\usepackage{graphicx}

\def\la{\mathrel{\mathchoice {\vcenter{\offinterlineskip\halign{\hfil
$\displaystyle##$\hfil\cr<\cr\sim\cr}}}
{\vcenter{\offinterlineskip\halign{\hfil$\textstyle##$\hfil\cr<\cr\sim\cr}}}
{\vcenter{\offinterlineskip\halign{\hfil$\scriptstyle##$\hfil\cr<\cr\sim\cr}}}
{\vcenter{\offinterlineskip\halign{\hfil$\scriptscriptstyle##$\hfil\cr<\cr
\sim\cr}}}}}

\def\ga{\mathrel{\mathchoice {\vcenter{\offinterlineskip\halign{\hfil
$\displaystyle##$\hfil\cr>\cr\sim\cr}}}
{\vcenter{\offinterlineskip\halign{\hfil$\textstyle##$\hfil\cr>\cr\sim\cr}}}
{\vcenter{\offinterlineskip\halign{\hfil$\scriptstyle##$\hfil\cr>\cr\sim\cr}}}
{\vcenter{\offinterlineskip\halign{\hfil$\scriptscriptstyle##$\hfil\cr>\cr
\sim\cr}}}}}

\title{Status of the KASCADE-Grande Experiment}

\author{K.-H. Kampert\address[uni]{%
  University of Karlsruhe, Institut f\"ur Kernphysik, P.O. Box 3640, 
  76021 Karlsruhe, Germany\\[-2.5mm]}
  \address[fzk]{%
  Institut\ f\"ur Kernphysik, Forschungszentrum Karlsruhe,
  76021~Karlsruhe, Germany\\[-2.5mm]}
  \thanks{Corresponding author: Karl-Heinz.Kampert@ik.fzk.de},
T.~Antoni\addressmark[uni], 
W.D.~Apel\addressmark[fzk],
F.~Badea\addressmark[uni]\thanks{on leave of absence from $\mathrm{^d}$},
K.~Bekk\addressmark[fzk], 
A.~Bercuci\addressmark[fzk]\footnotemark[2],
M. Bertaina\address[unito]{%
        Dipartimento di Fisica Generale dell'Universit\`a,
	10125 Torino, Italy\\[-2.5mm]},
H.~Bl\"umer\addressmark[fzk]\addressmark[uni],
H.~Bozdog\addressmark[fzk],
I.M.~Brancus\address[buk]{%
  National Institute of Physics and Nuclear Engineering,
  7690~Bucharest, Romania\\[-2.5mm]},
C.~B\"uttner\addressmark[uni],
A.~Chiavassa\addressmark[unito],
K.~Daumiller\addressmark[uni],
P.~Doll\addressmark[fzk],
J.~Engler\addressmark[fzk],
F.~Fe{\ss}ler\addressmark[uni],
P.L.~Ghia\address[cnrto]{%
        Istituto di Fisica dello Spazio Interplanetario del CNR,
	Sezione di Torino, 10133 Torino, Italy\\[-2.5mm]},
H.J.~Gils\addressmark[fzk],
R.~Glasstetter\addressmark[uni],
R.~Haeusler\addressmark[uni], 
A.~Haungs\addressmark[fzk], 
D.~Heck\addressmark[fzk], 
J.R.~H\"orandel\addressmark[uni],
A.~Iwan\addressmark[uni]\thanks{and University of Lodz, Poland}, 
H.O.~Klages\addressmark[fzk], 
G.~Maier\addressmark[fzk],
H.J.~Mathes\addressmark[fzk], 
H.J.~Mayer\addressmark[fzk], 
D.~Martello\addressmark[fzk]\thanks{permanent address: University 
Lecce, Italy}, 
J.~Milke\addressmark[uni], 
C.~Morello\addressmark[cnrto],
M.~M\"uller\addressmark[fzk],
G.~Navarra\addressmark[unito],
R.~Obenland\addressmark[fzk],
J.~Oehlschl\"ager\addressmark[fzk],
S.~Ostapchenko\addressmark[uni]\thanks{%
 on leave of absence from Moscow State University, Russia},
M.~Petcu\addressmark[buk], 
H.~Rebel\addressmark[fzk], 
M.~Roth\addressmark[fzk], 
H.~Schieler\addressmark[fzk], 
J.~Scholz\addressmark[fzk],
T.~Thouw\addressmark[fzk], 
G.C.~Trinchero\addressmark[cnrto],
H.~Ulrich\addressmark[uni],
R.~Ulrich\addressmark[uni],
J.H.~Weber\addressmark[uni], 
A.~Weindl\addressmark[fzk],
J.~Wentz\addressmark[fzk], 
J.~Wochele\addressmark[fzk], 
J.~Zabierowski\address[pol]
  {Soltan Institute for Nuclear Studies, 90950~Lodz, Poland},
and
S.~Zagromski\addressmark[fzk]
}
       
\begin{document}

\begin{abstract}
The status and capabilities of the KASCADE-Grande extensive air
shower experiment are presented.  The installation is located at
Forschungszentrum Karlsruhe and comprises a large collecting area
(0.5 km$^2$) electromagnetic array (Grande) operated jointly with
the existing KASCADE detectors.  KASCADE-Grande will cover the
primary energy range $10^{16}$ eV $< E_0 < 10^{18}$ eV
overlapping with KASCADE around $10^{16}$ eV, thus providing
continuous information on the primary energy and mass of cosmic
rays from $3 \cdot 10^{14}$ eV up to $10^{18}$ eV. The major
goal of the measurements is the unambiguous observation of
the``iron knee'' expected in the cosmic ray spectrum at $E_k^{Fe}
\approx 10^{17}$ eV.
%\vspace{1pc}
\end{abstract}

\maketitle

\section{INTRODUCTION}

The origin and acceleration mechanism of ultra-high energy cosmic
rays ($E \ga 10^{14}$ eV) and the origin of the `knee' at
approx.\ $3\cdot10^{15}$ eV have been subject to debate for
several decades.  Most commonly, it is assumed that cosmic rays
gain their energy by first order Fermi acceleration at highly
supersonic shock waves originating in supernova explosions. 
Simple dimensional estimates show that this process is limited to
$E_{\rm max} \la Z \times (\rho \times B)$, with $Z$ being the
atomic number of the cosmic ray (CR) isotope and $\rho$, $B$ the
size and magnetic field strength of the acceleration region.  A
more detailed examination of the astrophysical parameters
suggests an upper limit of acceleration of $E_{\rm max} \approx
Z\times 10^{15}$\,eV \cite{drury94b,berezhko99}.  It is thus
tempting to identify the knee with the maximum energy obtained by
such accelerators.  A brief compilation of alternative
interpretations of the knee is given in Ref.~\cite{kampert01}.  As
a simple consequence of the aforementioned `standard' picture,
one expects to see a break at constant rigidity for all particles,
i.e.\ protons would break off first and iron nuclei at
correspondingly higher energy.  This results in an increasingly
heavier composition at energies above the knee.

Such a global change of composition is in fact observed by
various experiments, including KASCADE
\cite{glasstetter99,antoni02} and EAS-TOP \cite{aglietta99} (a
recent compilation can be found in Ref.~\cite{swordy02}).  A
rather convincing proof has recently been presented by the
KASCADE collaboration \cite{ulrich01,kampert01} reporting
preliminary evidence for a knee shifting towards higher energies
with increasing mass of the primaries (see also contributions to
this conference by H. Ulrich {\it et al.} and M. Roth {\it et
al.}).  However, reconstruction properties and statistics do not
allow to clearly identify the expected iron knee at $E \approx
10^{17}$ eV. The primary goal of the KASCADE-Grande experiment is
to uncover the possible existence of an iron break by providing
high quality data from $10^{16}$ - $10^{18}$ eV. The experiment
covers an area of approx.\ 0.5~km$^2$ and is located next to the
KASCADE site in order to operate jointly with the existing
KASCADE detectors.  The combination with KASCADE will guarantee a
cross calibration of the detectors and will provide full coverage
from $3 \cdot 10^{14}$ to $10^{18}$ eV.

By the time of writing this article, all detector stations have
been deployed and first showers were already observed with
preliminary readout electronics.  Full operation will start in
January 2003 and full statistics will be reached after 3-4 years
of data taking.  The present status of the array, its resolutions
and capabilities in reconstructing the average primary
composition and its verification of the hadronic interaction
models are discussed.

\section{EXPERIMENT}

\begin{figure}[t]
\includegraphics[width=\columnwidth]{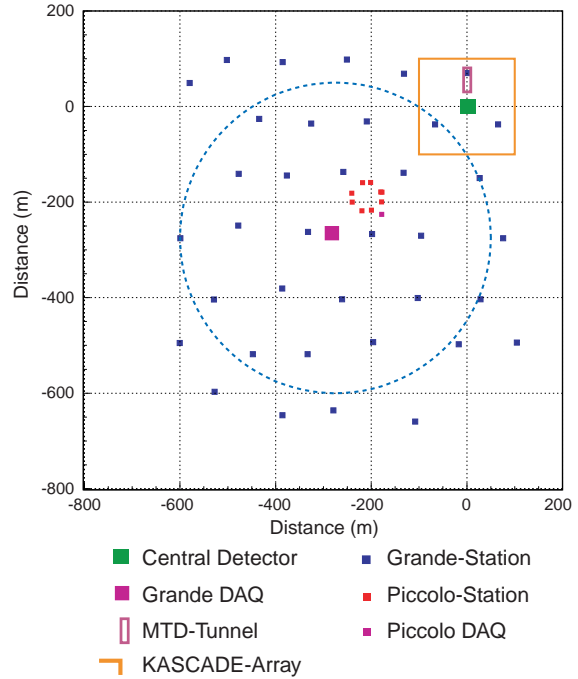}
\vspace{-1.0cm}
\caption{Layout of the KASCADE-Grande experiment. The circle 
indicates the acceptance region used in the analysis of 
Fig.~\ref{fig:trigeff}.}
\label{fig:layout}
\end{figure}

A sketch of the KASCADE-Grande layout is presented in Fig. 
\ref{fig:layout}.  It consists of three different detector
arrays: KASCADE, Grande, and Piccolo.

%\begin{figure}[!ht]

The KASCADE experiment \cite{kascade} comprises three major
components: an array of electron and muon detectors, a central
detector mainly for hadron measurements but with substantial muon
detection areas, and an underground muon tracking detector (MTD). 
The $e/\gamma$- and $\mu$-detectors cover an area of 200 $\times$
200 m$^2$ and consist of 252 detector stations located on a
square grid of 13 m separation providing in total about 490 m$^2$
of $e/\gamma$ and 622 m$^2$ of muon coverage.  The detection
thresholds for vertical incidence are $E_e > 5$ MeV and $E_{\mu}
> 230$ MeV. The central detector system (320 m$^2$) consists of a
highly-segmented hadronic calorimeter read out by 44,000 channels
of warm liquid ionization chambers, a layer of scintillation
counters above the shielding, a trigger plane of scintillation
counters in the third layer and, at the very bottom, 2 layers of
positional sensitive MWPC's and a streamer tube layer with pad
read-out for muon tracking at $E_{\mu} \ge 2.4$ GeV. The MTD is
located in a 44 $\times$ 5.4 $\times$ 2.4 m$^3$ tunnel, close to
the central detector.  It houses three horizontal and two
vertical layers of positional sensitive limited streamer tubes
for muon tracking at $E_{\mu} \ge 0.8$ GeV. The accuracy in
reconstructing the muon direction is $\sigma \approx 0.35^\circ$ 
\cite{doll02}. 
With its approx.\ 1000~m$^{2}$ of muon detector coverage
KASCADE-Grande provides a much better muon sampling fraction
(approx.\ 1000 m$^{2}$) than any other previous experiment.
 
Grande comprises 37 stations of 10 m$^2$ scintillation counters
reused from the former EAS-TOP experiment \cite{as89}.  The
stations are positioned at an average mutual distance of approx.\
130~m.  They are subdivided into 16 individual scintillation
detectors (80 $\times$ 80 cm$^{2}$) read out by one phototube
each.  The inner four scintillation detectors are independently
read out by low gain phototubes in order to provide a high
dynamic range with high quality particle density and timing
measurements from $\approx 0.3$ to $\approx 30000$ charged
particles per 10~m$^{2}$ \cite{bertaina01}.

\begin{figure}[t]
\includegraphics[width=\columnwidth]{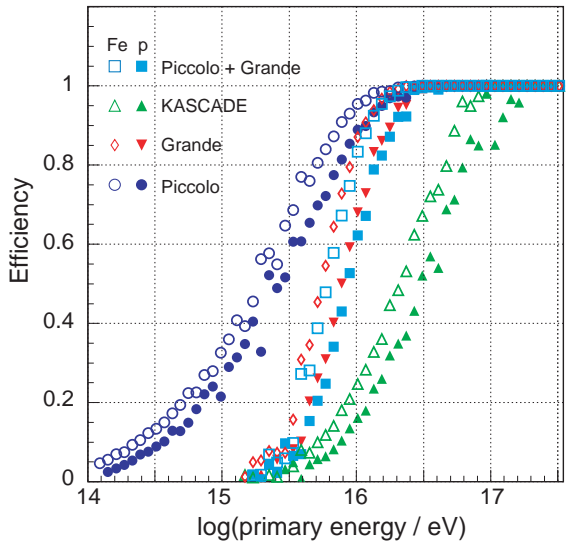}
\vspace{-1cm} \caption{Simulated trigger efficiency for
KASCADE-Grande.  The up-triangles show the efficiency of the
KASCADE array to trigger also the readout of Grande, the
down-triangles and diamonds the efficiency of Grande (7-fold)
only, the boxes show the efficiency that Piccolo is triggered
together with Grande, and the circles show the efficiency for
Piccolo to be triggered by showers in the acceptance region of
Fig.~\ref{fig:layout}.}
\label{fig:trigeff}
\end{figure}

Piccolo consists of an array of 8 stations, each equipped with 12
scintillator plates (10 m$^2$/station) that were reused from the
KARMEN neutrino experiment \cite{karmen}.  The small array is
placed between the centres of the KASCADE and Grande arrays and
its main task is to provide a fast external trigger to Grande and
KASCADE allowing to record coincident events with all the
detectors of KASCADE-Grande.  The simulated trigger efficiency
using CORSIKA/QGSJET calculations is shown in
Fig.~\ref{fig:trigeff} together with the trigger efficiency of
the Grande array only.  Using Piccolo, full detection efficiency
is thus reached for $E \ga 10^{16}$ eV while in its absence, the
KASCADE multidetector information would only be availabe for
showers above energies of $10^{17}$ eV. A summary of the detector
components together with their most relevant parameters is given
in Table \ref{tab:compon}.

For triggering, Grande is electronically subdivided into 18
hexagonal clusters of 6 outer and one central station each. 
Requiring e.g.\ 4-fold coincidences within each cluster
(demanding a signal over threshold in the central and in 3
adjacent outer stations) we measure a trigger rate of $\sim 0.3$
Hz per cluster and a total trigger rate of Grande of $\sim 5.9$
Hz, well in expectation with simulation results.  To improve the
signal/noise ratio for low particle densities, the PMT signals
will be amplified and shaped in the Grande stations. 
Digitisation is done after transmission via 700~m long shielded
cables in peak sensing ADCs located in the central Grande-DAQ
station.  All detector components are deployed and data taking
will start in January 2003.

\begin{table*}[th]
\begin{center}
\begin{tabular}{cccccc}
\hline
Detector & \# of Channels & Spacing (m) & Total Area ($m^2$) & Threshold
$E_{kin}$  & Particle Type\\
\hline
Array $e/ \gamma$ & 252 & 13 & 490 & 5 MeV & $e$\\
Array $\mu$       & 192 & 13 & 622 & 230 MeV $\times \sec \theta$ & $\mu$\\
Trigger Plane     & 456 &  - & 208 & 490 MeV $\times \sec \theta$ & $\mu$\\
MWPCs           & 26080 &  - & 129 & 2.4 GeV $\times \sec \theta$ & $\mu$\\
Calorimeter     & 44000 &  - & 304 & 50  GeV & Hadrons \\
Muon Tunnel     & 24576 &  - & 128 & 800 MeV & $\mu$\\
Grande          &    74 &$\langle$130$\rangle$ & 370 & 3 MeV   & $e$,$\mu$\\
Piccolo         &    16 &$\langle$25$\rangle$ &  80 & 5 MeV   & $e$,$\mu$\\
\hline
\end{tabular}
\end{center}
\vspace{-0.1cm} \caption{Compilation of detector components and
their particle detection thresholds.}
\label{tab:compon}
\end{table*}

\section{ANALYSIS}

After three years of data taking the total exposure of
KASCADE-Grande will be $\Gamma \simeq 10^{14}$ m$^2$\,sr\,s. 
This corresponds to the following numbers of collected events
(including the trigger efficiency of Piccolo):
$n(>10^{16} eV) \simeq 1.7 \cdot 10^6$,
%$n(>3 \cdot 10^{16} eV) \simeq 2.8 \cdot 10^5$,
$n(>10^{17} eV) \simeq 2.5 \cdot 10^4$,
%$n(>3 \cdot 10^{17} eV) \simeq 2.8 \cdot 10^3$,
$n(>10^{18} eV) \simeq 250$. 
KASCADE-Grande will thus provide statistically significant
physical information up to about $10^{18}$~eV.

\begin{figure}[t]
\includegraphics[width=\columnwidth]{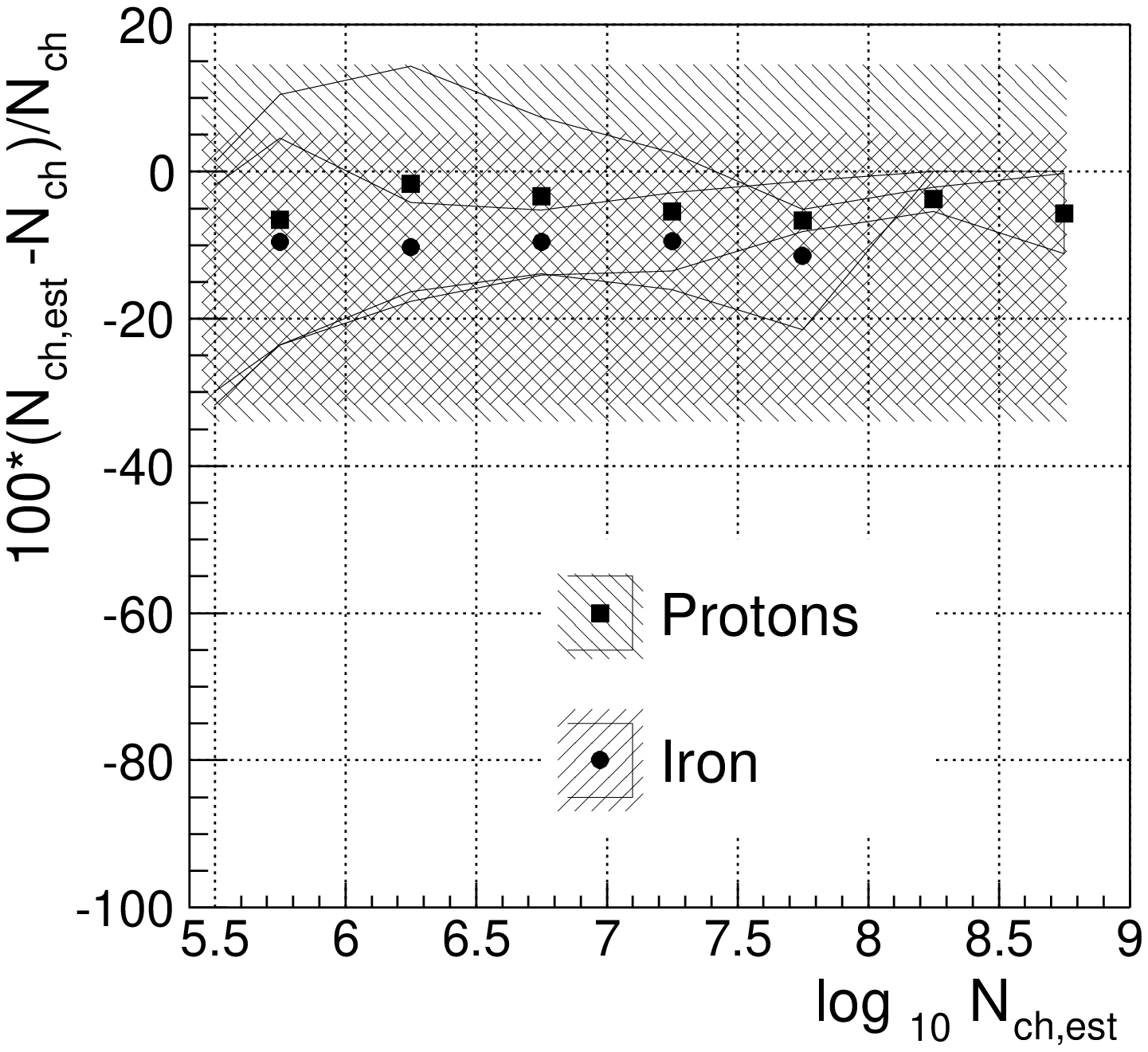}
\vspace{-1cm} \caption{Simulated resolution of the
reconstructed shower size in Grande for proton and iron induced
primaries using CORSIKA-QGSJET simulations and requiring 7-fold 
coincidences.}
\label{fig:nresol}
\end{figure}

The basic experimental observables will be: (i)~the muon density
at core distances between 300 and 600~m providing, after applying
a fit to the observed muon lateral distribution, the muon density
at 600~m, $\rho_{\mu600}$, (ii) the muon production heights
($h_\mu$) reconstructed from the KASCADE muon tracking detector
by means of triangulation (including timing information for the
muons in the KASCADE central detector), and (iii) the shower size
($N_{ch}$) and lateral electron density profile from the extended
electromagnetic (e.m.) Grande array.

The reconstruction accuracies of the Grande array have been
estimated by CORSIKA-QGSJET simulations.  For the energy range of
interest ($N_{ch} \ge 10^{7}$; $E \ga 10^{16}$ eV, vertical
showers) the shower core position can be reconstructed to better
than 5~m and the shower direction to better than $1.5^{\circ}$. 
Using not yet optimized lateral distribution functions, the
shower size can be reconstructed to better than 10\,\% (see
Fig.~\ref{fig:nresol}).

Typical muon densities for $E_0 = 10^{17}$eV protons will be
$\rho_{\mu 600} \approx 0.1$ m$^{-2}$.  Considering a detection
area of about 800 m$^2$ for low energy muons, we expect an
average of 80 muons detected on a event by event basis meaning a
statistical fluctuation of about $12\%$.

\begin{figure}[t]
\includegraphics[width=8.0cm,height=6.6cm]{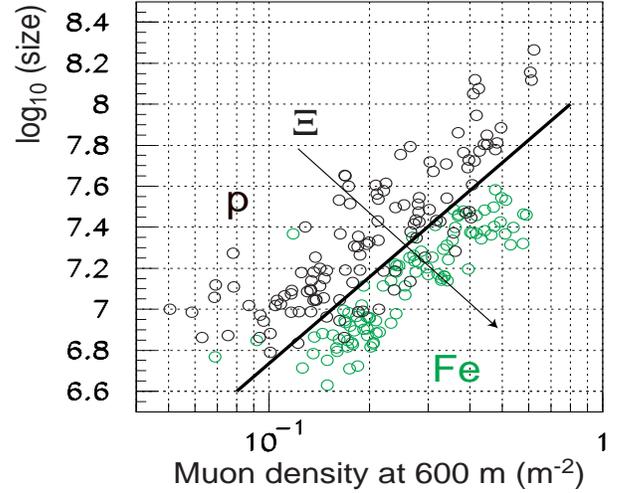} 
\vspace{-1cm} \caption{QGSJET Simulations of proton and iron
primaries in the energy range of $2 \cdot 10^{16} \le E \le 5
\cdot 10^{17}$ eV. The light symbols represent iron and the dark
ones proton simulations.  Both, shower fluctuations and
experimental uncertainties are taken into account.  The $\Xi$
line is used as a guideline for the $\Xi$ mass axis.}
\label{fig:ximass}
\end{figure}

Monte Carlo simulations using the CORSIKA package have been
performed also to check the experiment capabilities for testing
the hadronic interaction models and reconstructing the mass of
primary CRs through the three observables $N_e$, $\rho_{\mu
600}$, and $h_\mu$.  An example is presented in
Fig.~\ref{fig:ximass}.  Here, a $N_e$ vs $\rho_{\mu 600}$ scatter
plot is shown for proton and iron induced simulated air showers. 
The primary composition is simply defined in terms of the mass
parameter $\Xi$, whose definition through $N_e$ and $\rho_{\mu
600}$ is shown in the same figure.  The line separating p and Fe
primaries best is indicated by the bold line.  The mass axis
$\Xi$ is thus shown orthogonal to this line.  The parameter
$\rho_{\mu 600}$ can, as first approximation, be used as energy
estimator.  It has been verified that the composition can be
reconstructed in a reliable way even by such simple methods. 
However, the later analysis will use more advanced techniques,
similar to those of Refs.~\cite{ulrich01,antoni02b}.

Hadronic interaction models will be studied by their differences
seen in the longitudinal shower developments.  Experimentally,
this will be performed by the observed average relation between
$h_{\mu}$ and $\rho_{\mu 600}$ and, to some extend, by studying
hadronic observables using the KASCADE calorimeter.

\section{SUMMARY AND CONCLUSIONS}

KASCADE-Grande has been realized at the Forschungszentrum
Karlsruhe by a joint operation of the KASCADE and EAS-TOP
detectors.  The apparatus is almost completed and data taking
will start shortly.  Within three years of effective data taking
it will accumulate sufficient statistics up to $10^{18}$ eV. The
experiment will cover an energy range that is only poorly known,
essentially from old AKENO \cite{as84} and the lower tail of
FLY'S EYE \cite{as94} data.  Its task is to provide more
information on the structure of the knee by testing the rigidity
dependent model up to the energies of the ``iron'' group. 
Moreover, it will allow to test hadronic models in an energy
range not accessible to present accelerators but important for
CRs physics.  Finally, it will provide a bridge to the
measurement and interpretation of CRs for experiments working at
much higher energies like the Pierre Auger, HIRES or EUSO
projects.

\subsection*{Acknowledgements}
The authors are indebted to the members of the engineering and
technical staff of the KASCADE-Grande collaboration, who
contributed with enthusiasm and engagement to the construction of
the experiment.  The work has been supported by the Ministry for
Research and Education of the Federal Government of Germany. 
Special thanks also go to INFN for allowing the use of the
EAS-TOP equipment to build the Grande array.  The collaborating
Institute of Lodz is supported by the Polish State Committee for
Scientific Research and the Institute of Bucharest by a grant of
the Romanian National Academy for Science, Research and
Technology.  The KASCADE-Grande work is embedded in the frame of
scientific -technical cooperation (WTZ) projects between Germany
and Romania, and between Germany and Poland.

\end{document}